\documentclass[secnumarabic, graphics,floatfix, nofootinbib,tightenlines,nobibnotes, aps, prl, 12pt]{revtex4-1}
\usepackage{graphicx}
\usepackage[english]{babel}
\usepackage[utf8]{inputenc}
\usepackage{amsmath,amssymb}
\usepackage{amsfonts}
\usepackage{multirow}
\usepackage{pstricks}
\usepackage{float}
\usepackage{subfigure}
\usepackage{color}
\usepackage{epsfig}
\usepackage{pst-tools}

\begin{document}

\title{A quasi-particle model with a phenomenological critical point}
\author{Hong-Hao Ma$^1$}
\author{Danuce Marcele Dudek$^2$}
\author{Kai Lin$^{3}$}
\author{Wei-Liang Qian$^{4,1,5}$\footnote{Presentation given at XIV International Workshop on Hadron Physics, 18-23 March, 2018, Florian\'opolis, SC, Brazil}}
\author{Yogiro Hama$^{6}$}
\author{Takeshi Kodama$^{7,8}$}

\affiliation{$^1$ Faculdade de Engenharia de Guaratinguet\'a, Universidade Estadual Paulista, 12516-410, Guaratinguet\'a, SP, Brazil}
\affiliation{$^2$ Universidade Federal Fronteira Sul, Campus Realeza, 85770-000, Realeza, PR, Brazil}
\affiliation{$^3$ Instituto de F\'isica e Qu\'imica, Universidade Federal de Itajub\'a, 37500-903, Itajub\'a, Brazil}
\affiliation{$^4$ Escola de Engenharia de Lorena, Universidade de S\~ao Paulo, 12602-810, Lorena, SP, Brazil}
\affiliation{$^5$ School of Physical Science and Tecnology, Yangzhou University, 225002, Yangzhou, Jiangsu, P.R. China}
\affiliation{$^6$ Instituto de F\'isica, Universidade de S\~ao Paulo, C.P. 66318, 05315-970, S\~ao Paulo-SP, Brazil}
\affiliation{$^7$ Instituto de F\'isica, Universidade Federal do Rio de Janeiro, C.P. 68528, 21945-970, Rio de Janeiro-RJ , Brazil}
\affiliation{$^8$ Instituto de F\'isica, Universidade Federal Fluminense, 24210-346, Niter\'oi-RJ, Brazil}

\date{May 3rd, 2018}

\begin{abstract}
A hybrid parameterization of a quasiparticle equation of state is proposed, with a critical point implemented phenomenologically.
On the one hand, a quasiparticle model with finite chemical potential is employed for the quark-gluon plasma phase, calibrated to the lattice quantum chromodynamics data.
On the other hand, the low-temperature region for the hadronic phase of the matter is described by the hadronic resonance gas model with excluded volume correction.
A particular interpolation scheme is adopted so that the phase transition is a smooth crossover for small chemical potential.
A phenomenological critical point is implemented beyond which the phase transition becomes that of the first order.

\pacs{12.38.Bx, 12.38.Aw, 11.15.Bt}

\end{abstract}

\maketitle

\section{Introduction}

An interesting phenomenological approach to address the thermodynamic properties of quark-gluon plasma (QGP) is the so-called quasiparticle model.
The model provides an intuitive interpretation for the thermodynamical properties of the system obtained by lattice quantum chromodynamics (QCD) simulations.
The latter corresponds to the region of transition where $T\sim T_c$.
It is understood to possess a dramatic change of the number of degrees of freedom, as the system is dominated by intrinsically nonperturbative interactions.
Inspired by its counterparts in other fields of physics, the quasiparticle model views the strongly interacting matter as consisting of non-interacting quanta, which carry the same quantum numbers of quarks and gluons.
Subsequently, the strong and nonperturbative interactions among the elementary constituents are incorporated through the medium dependent mass.
The model was first proposed by Peshier $et~al.$~\cite{Peshier:1994zf}, and then reformulated by Gorenstein and Yang~\cite{Gorenstein:1995vm} by addressing the issue of thermodynamical consistency.
Thereafter, many alternative approaches have been discussed~\cite{Biro:2001ug,Bannur:2006hp,Bannur:2007tk,Chandra:2011en,Oliva:2013zda}.
In the Refs.~\cite{Bannur:2006hp,Bannur:2007tk}, it is assumed that the form of the internal energy and the particle number remains unchanged as in statistical mechanics.
On the other hand, the pressure and additional thermodynamic quantities are derived.
The formula is claimed to be consistent with the thermodynamic relation.
Lattice QCD studies~\cite{lattice-01,lattice-02} showed that the transition is a crossover at vanishing baryon density.
At non-vanishing chemical potential, a first order phase transition is indicated by various models~\cite{Halasz:1998qr,Berges:1998rc,Stephanov:1998dy}.
These results imply that the phase diagram is probably featured by a critical point at which the line of first-order phase transitions ends.
In fact, the existence of such a critical point is still being debated.

Recently, we revisited the thermodynamical consistency and proposed a hybrid equation of state (EoS)~\cite{Ma:2017kzz,Ma:2018bwf} which can be used by the hydrodynamical models for relativistic heavy ion collisions.
In our model, the EoS connects QGP phase to the hadronic phase where a phenomenological critical point is implemented according to the scheme proposed by Hama $et~al.$~\cite{Hama:2005dz}.
For the QGP phase, a quasiparticle model with finite chemical potential is used~\cite{Bannur:2006hp}, adjusted to the recent Lattice QCD results~\cite{Borsanyi:2012cr,Borsanyi:2013bia}.
The hadronic phase is described by a hadronic resonance gas (HRG) model with excluded volume correction~\cite{Rischke:1991ke,Hama:2004rr}.
The critical point is implemented phenomenologically at finite baryon chemical potential.
More specific details of the model can be found in~\cite{Ma:2017kzz,Ma:2018bwf}.

\section{The formalism}

To reproduce the lattice QCD data for 2+1 flavor QGP at high temperature, we employ the quasiparticle model proposed in~\cite{Bannur:2006hp}.
The approach keeps the form of energy and particle number the same as formulated as grand ensemble averages in statistical mechanics.
The system is viewed as consisting of a collection of non-interacting quasiparticles of the gluons, up, down as well as strange quarks.
The single particle energy of quasiparticles $\omega_k$ satisfies the on-shell dispersion relation
\begin{eqnarray}
\omega_k^2=k^2+m_{g,q}^2  ,
\end{eqnarray}
where for the thermal masses of quasiparticles, one adopts the following prescription
\begin{eqnarray}
m_{g}^{2}=\frac{3}{2} \omega_{p}^{2}
\end{eqnarray}
for gluons and
\begin{eqnarray}
m_{q}^{2}=(m_{q0}+m_f)^2 + m_{f}^{2}
\end{eqnarray}
for quarks, where $q$ stands for $u$, $d$, or $s$ quark.
Here $m_{q0}$ stand for the current mass of the quarks.
We take $m_{s0}=0.150$ GeV for strange quark, and $m_{u0,d0}=m_{s0}$/28.15 $\approx$ 5.33 MeV for up and down quarks.
The plasmon frequency $\omega_p$ and the effective mass of soft massless quark $m_f$ are associated with the collective behavior of the system.
For zero chemical potential, we adopt the parameterization of model II proposed in \cite{Bannur:2007tk} inspired by the resummed hard thermal loop (HTL) approximation as follows
\begin{eqnarray}
\omega_{p}^{2}=&&a_{g}^{2} g^2 \frac{n_g}{T} + \sum_{q} a_{q}^{2} g^2 \frac{n_q}{T}, \label{vmbmass1}\\
m_{f}^{2}= &&b_{q}^{2} g^2 \frac{n_q}{T}, \label{vmbmass2}
\end{eqnarray}
where $n_g$ and $n_q$ are number densities of gluons and quarks.
Here the coefficients $a_g$, $a_q$ and $b_q$ are to be determined by demanding Eqs.(\ref{vmbmass1}-\ref{vmbmass2}) approach the perturbative results as $T \rightarrow \infty$.

The principle of asymptotic freedom indicates that the effective coupling constant $g$ falls with increasing temperature.
The coupling constant can be obtained for finite temperature up to two-loop approximation~\cite{Caswell:1974gg,BeiglboCk:2006lfa} and generalized to the case of finite chemical potential following ~\cite{Bannur:2007tk}.
Also, at finite baryon density, the plasma frequencies are taken to be~\cite{Peshier:1999ww}
\begin{eqnarray}
m_{f}^{2}=&& \frac{g^2 T^2}{18}n_f (1+\frac{\mu^{2}}{\pi^2 T^2}) \label{mfmu}.
\end{eqnarray}
where $n_f$ is the number of flavors.
For the application of relativistic heavy ion collisions, we consider strangeness neutrality condition.
The above system of coupled equations thus can be solved self-consistently for plasma frequency and number density.

For the description of HRG model, the pressure is dertermined by the following self-consistent equations~\cite{Rischke:1991ke}
\begin{eqnarray}
p^{H}(T,\mu_B,\mu_S,\mu_3)= \sum_{i=1} p_{i}^{id}(T,\widetilde{\mu_{i}}) ,\label{hrgeq}\\
\widetilde{\mu_{i}} \equiv  \mu_i - v_i p^{H} .\nonumber
\end{eqnarray}
where the excluded volume $v_{i}=(4 \pi r_{0}^{3}/3)$, with $r_{0}=0.7fm$ for baryons and $r_{0}=0$ for mesons.
If the phase transition is of the first order, the chemical potential and temperature of the two phases are determined by the Gibbs condition.
In order to describe a smooth crossover in the region of small baryon density, we adopt the following scheme~\cite{Hama:2005dz}
\begin{eqnarray} \label{Gibbs}
(p-p^Q)(p-p^H)=\delta(\mu,T) ,
\end{eqnarray}
where
\begin{eqnarray}
\delta(\mu,T)=\delta_{0}(T) \exp \left[-(\mu/\mu_c)^4\right] ,
\end{eqnarray}
and $\mu_c$ is the critical chemical potential, which is taken to be $\mu_c = 0.3$ GeV in this work.
We note, when $\delta_0=0$, a first order phase transition is recovered.
Eq.(\ref{Gibbs}) can be solved straightforwardly, and one subsequently obtains the expressions for entropy density, baryon density, and energy density.
We choose $\delta_{0}(T)$ to be a piecewise function as follows
\begin{itemize}
  \item $\delta_{0}(T)=\delta_{0} e^{-c (T-T_{p})^{2}}, ~~~~~~~~T \leq T_{p}$
  \item $\delta_{0}(T)=\delta_{0},~~~~~~~~~~~~~~~~~~~~T_{p}<T \leq T_{p}+0.02$
  \item $\delta_{0}(T)=\delta_{0} e^{-c (T-T_{p}-0.02)^{2}},~~T > T_{p}+0.02$
\end{itemize}
where $\delta_{0}=5.90 \times 10^{-10}$ GeV$^8$ and $c=10^{3}$.
$T_{p}$ stands for the temperature (in GeV) of the corresponding first order transition.

\section{Numerical Results}

\begin{figure}
\begin{tabular}{cc}
\begin{minipage}{250pt}
\centerline{\includegraphics[width=250pt]{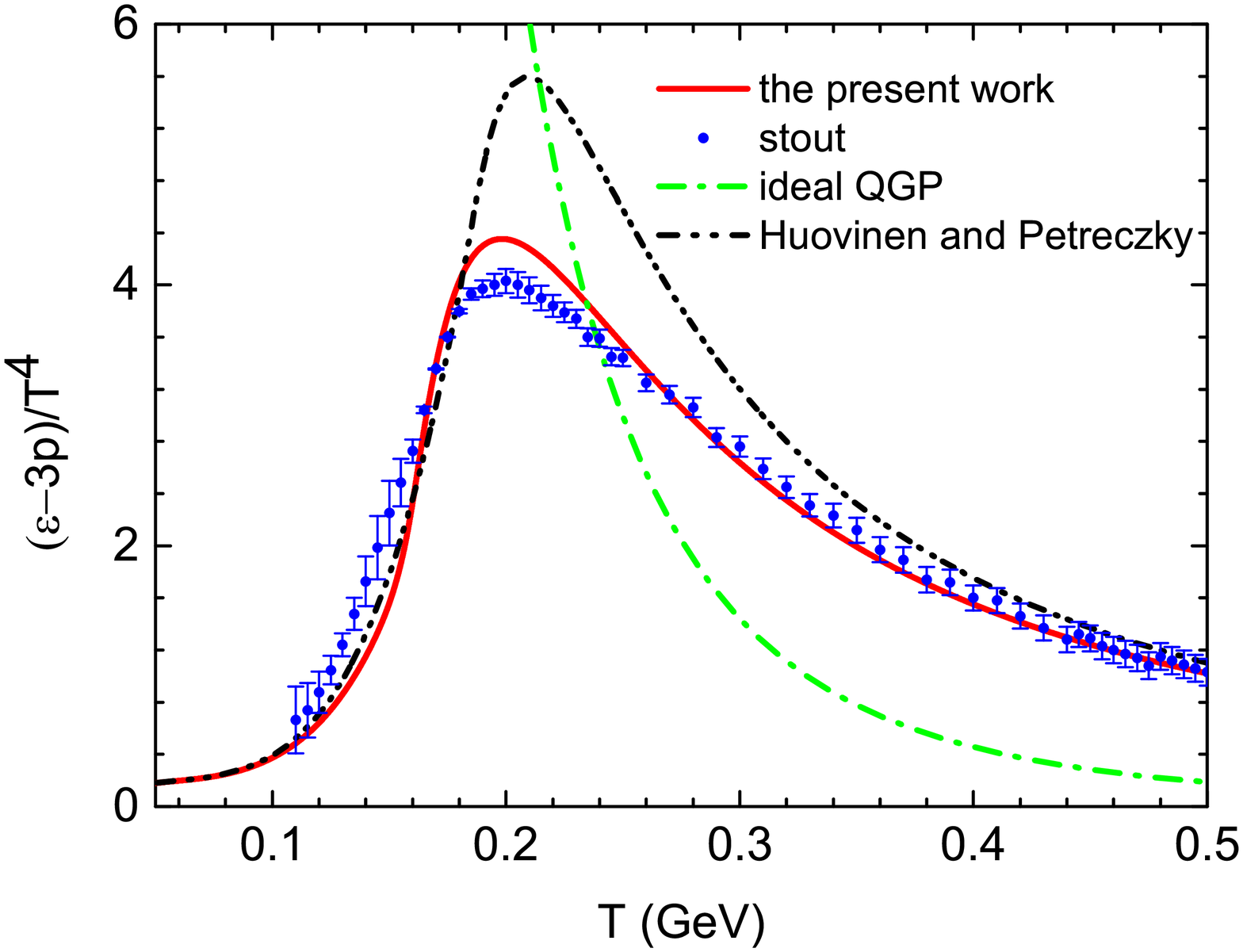}}
\end{minipage}
&
\begin{minipage}{250pt}
\centerline{\includegraphics[width=250pt]{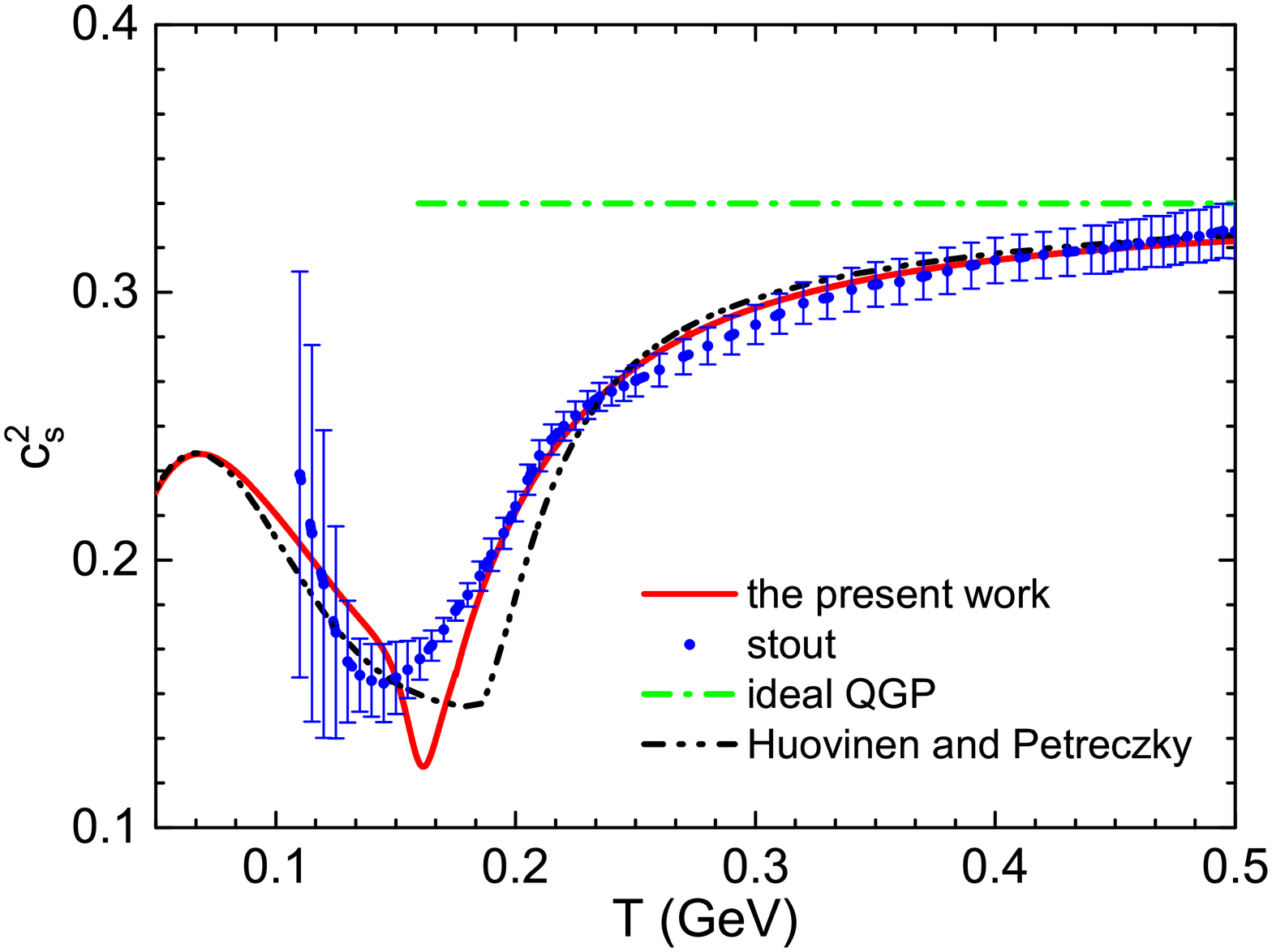}}
\end{minipage}
\\
\begin{minipage}{250pt}
\centerline{\includegraphics[width=250pt]{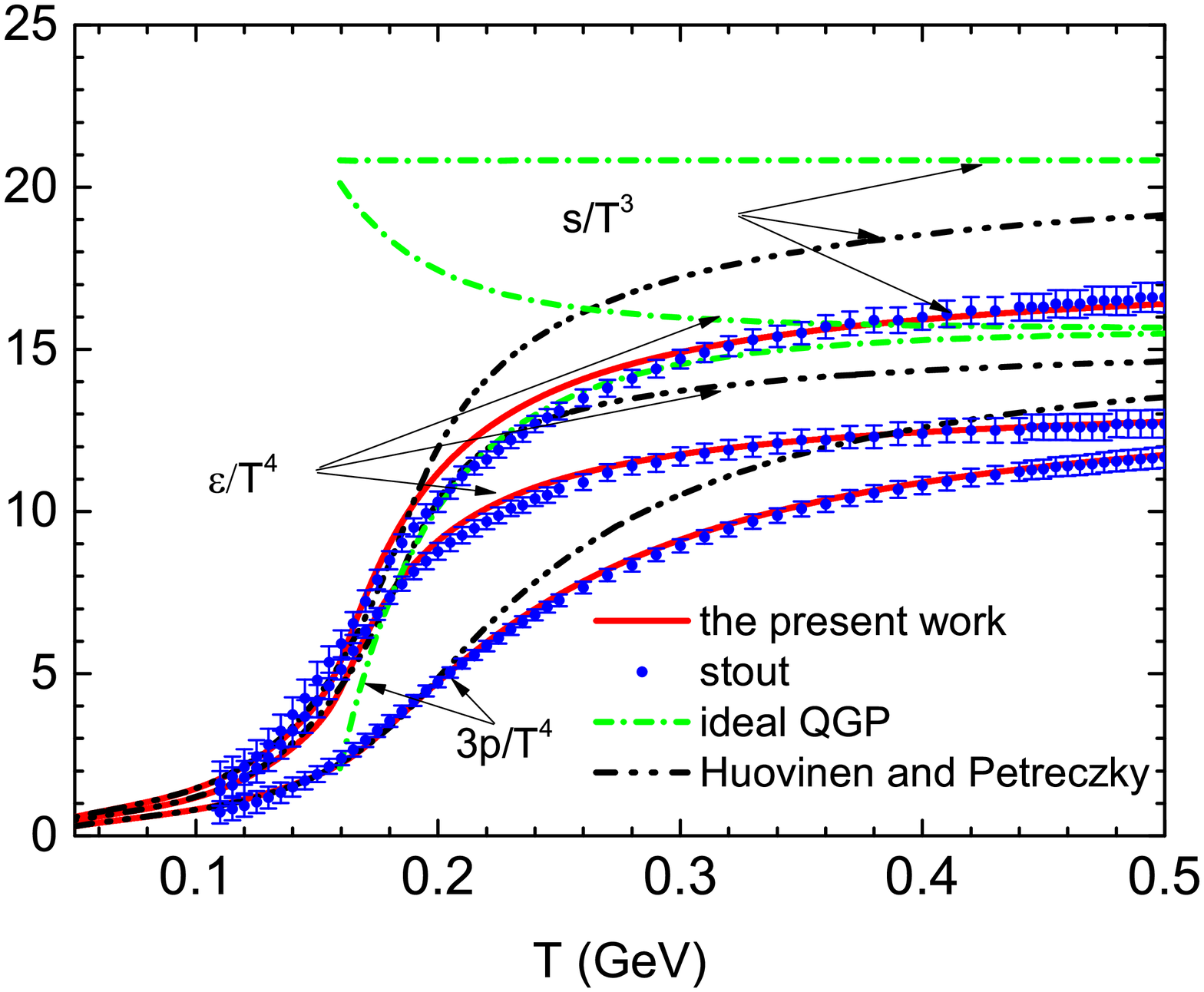}}
\end{minipage}
&
\begin{minipage}{250pt}
\centerline{\includegraphics[width=250pt]{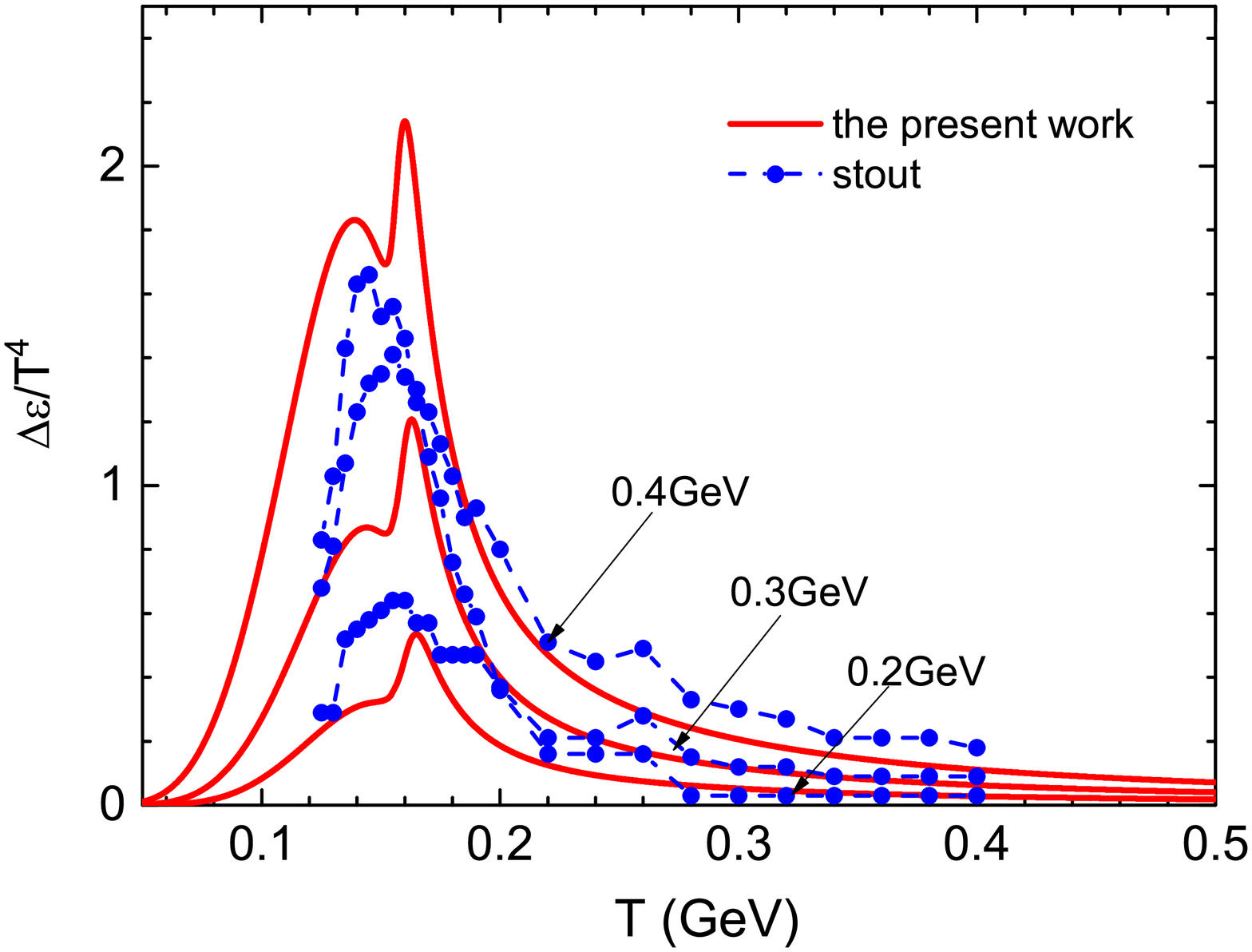}}
\end{minipage}
\\
\end{tabular}
 \caption{(Color online) The calculated results of the trace anomaly, speed of sound, pressure, entropy and energy density in comparison with the lattice QCD data~\cite{Borsanyi:2012cr,Borsanyi:2013bia}, the EoS obtained by Huovinen and Petreczky in Ref~\cite{eos-pasi-02} and bag model of QGP.
The calculated results of the present model are represented by red solid curves; lattice QCD simulations by stout action (indicated by ``stout") are shown in dotted blue curves with uncertainties; those obtained by Huovinen and Petreczky (indicated by ``Huovinen and Petreczky") are depicted by black dash-dot-dot curves and the bag model (indicated by ``ideal QGP") in dash-dot curves.}
 \label{tracecs2}
\end{figure}

Here we present the numerical results of the obtained EoS.
As in \cite{Schneider:2001nf,Ivanov:2004gq}, an overall normalization factor 1.06 is introduced to take into account the unknown correction to the effective number of degrees of freedom.
For zero chemical potential, the resulting entropy density, energy density, and pressure are presented in the left plot of the second row of Fig.\ref{tracecs2} in comparison with results of the lattice QCD~\cite{Borsanyi:2012cr,Borsanyi:2013bia}.
As a comparison, we also depicted the results of a well-known parameterized EoS obtained by Huovinen and Petreczky~\cite{eos-pasi-02} and those of the bag model.
We see that, for all three quantities, the recent lattice data are reasonably well reproduced.
The difference between present results and those by Huovinen and Petreczky~\cite{eos-pasi-02} is because the EoS was fit to the earlier lattice data, which has been improved continuously in time.
The bag model serves to demonstrate the significant difference between the lattice data and those of the ideal gas model.
For instance, the function $\varepsilon/T^4$ vs. $T$ is a decreasing function while lattice calculations show that it monotonically increases with temperature $T$.
Also, the strong interaction causes distinct deviation from Boltzmann limit as presented by the difference for $\varepsilon/T^4$ at moderate temperature $T\sim 0.5$ GeV. 
The calculated trace anomaly and the speed of sound are presented in the first row of Fig.\ref{tracecs2}.
It is found that the trace anomaly is reasonably well reproduced, the maximum of the curve is near $T\sim 0.2$ GeV.
In this region, the present model reproduces the pressure well in this region but slightly overestimates the energy density. 
Moreover, the deviation of $\varepsilon$ from the lattice data increases with increasing temperature in the vicinity of $T\sim 0.2$ GeV.
As a result, the maximum of the calculated trace anomaly overestimate the lattice data and is slightly shifted towards the right.
For the bag model, the trace anomaly diverges as $T\rightarrow 0$, instead of being identically zero, which is due to the existence of the bag constant.
As shown in the plot on the right of the first row, the main feature of the speed of sound is also obtained.
At high temperature, one finds that the speed of sound approaches that of the ideal gas.
As $T$ decreases, the speed of sound decreases and reaches a minimum. 
Comparing to the lattice results, the location of the minimum is slightly shifted towards higher temperature.
Since the speed of sound is related to the ratio of the derivatives of energy density and pressure, the result turns out to be more sensitive to the choice of parameterization.
To be specific, in the region $T\sim 0.15$ GeV, the derivative $d\varepsilon/dT$ slightly underestimates the data at low temperature, namely, the calculated curve $\varepsilon/T^4$ is a bit too flat comparing to the data and then it becomes steeper as the temperature increases, while $dp/dT$ behaves oppositely in this region.
Consequently, the calculated sound speed underestimates the lattice data and the minimum is slightly shifted to the right.
Since the properties of the system at $T\sim 0.15$ GeV is mostly determined by the HRG model, one observes that the use of a fine-tuned model might further improve the result.
For finite chemical potential, differences in pressure, energy density, particle number density can be calculated.
The results for energy density are shown in the right plot of the second row of Fig.\ref{tracecs2}, in comparison with the lattice QCD results by stout action~\cite{Borsanyi:2012cr}.
Though the lattice QCD results are qualitatively reproduced, it is found that the region connecting the two phases possesses a secondary peak, which is probably related to the first order phase transition at large baryon density in the model and the adopted smoothing parameterization.

\section{Concluding remarks}

To summarize, an interpolation scheme is adopted to build an EoS with a phenomenological critical point at finite chemical potential.
A quasiparticle model is fitted to the lattice QCD data to describe the high-temperature QGP phase, while an HRG model with exclusive volume correction is utilized for the hadronic phase in the low-temperature region.
The critical point is implemented so that all other quantities are derived from the Gibbs thermodynamic potential.
As the EoS plays an essential role in the hydrodynamic description of relativistic heavy-ion collisions~\cite{sph-review-1}, we plan to carry out a hydrodynamic study of the relevant quantities associated with the Beam Energy Scan program of RHIC using the present EoS in the near future.

\section*{Acknowledgments}
We gratefully acknowledge the financial support from FAPESP, FAPERJ, CNPq and CAPES.

\bibliographystyle{h-physrev}
\bibliography{references_ma,references_qian}

\end{document}